\documentclass[%
 prx,
rsi,%
 amsmath,amssymb,
 reprint,%
superscriptaddress
]{revtex4-1}

\usepackage{graphicx}
\usepackage{dcolumn}
\usepackage{bm}
\usepackage{color}

\begin{document}

\title{Dynein catch bond as a mediator of codependent bidirectional cellular transport}%

\author{Palka Puri}
\affiliation{Department of Physics, IIT Bombay, Mumbai, India}
\affiliation{Institute for Nonlinear Dynamics, Georg August University of Goettingen, Germany}
\author{Nisha Gupta}
\affiliation{Indian Institute of Science Education and Research, Mohali, India}
\author{Sameep Chandel}
\affiliation{Indian Institute of Science Education and Research, Mohali, India}
\author{Supriyo Naskar}
\affiliation{Department of Physics, Indian Institute of Science, Bangalore, India}
\author{Anil Nair}
\affiliation{Department of Physics, Savitribai Phule Pune University, Pune, India}
\author{Abhishek Chaudhuri}
\email{abhishek@iisermohali.ac.in }
\affiliation{Indian Institute of Science Education and Research, Mohali, India}
\author{Mithun K. Mitra}
\email{mithun@phy.iitb.ac.in}
\affiliation{Department of Physics, IIT Bombay, Mumbai, India}
\author{Sudipto Muhuri}
\email{sudipto@physics.unipune.ac.in}
\affiliation{Department of Physics, Savitribai Phule Pune University, Pune, India}

\date{\today}

\begin{abstract}
Intracellular bidirectional transport of cargo on microtubule filaments is achieved by the collective action of oppositely directed dynein and kinesin motors.  Experiments have found that in certain cases, inhibiting the activity of one type of motor results in an overall decline in the motility of the cellular cargo in both directions. This counter-intuitive observation,  referred to as {\em paradox of codependence} is inconsistent with the existing paradigm of a mechanistic tug-of-war between oppositely directed motors. Unlike kinesin, dynein motors exhibit catchbonding, wherein the unbinding rates of these motors decrease with increasing force on them.  Incorporating this catchbonding behavior of dynein in a theoretical model, we show that the functional divergence of the two motors species manifests itself as an internal regulatory mechanism, and leads to codependent transport behaviour in biologically relevant regimes. Using analytical methods and stochastic simulations, we analyse the processivity characteristics and probability distribution of run times and pause times of transported cellular cargoes. We show that catchbonding can drastically alter the transport characteristics and also provide a plausible resolution of the paradox of codependence.
\end{abstract}

\maketitle

\section{Introduction}

\begin{figure*}[t!]
    \centering
    \includegraphics[width=0.8\linewidth]{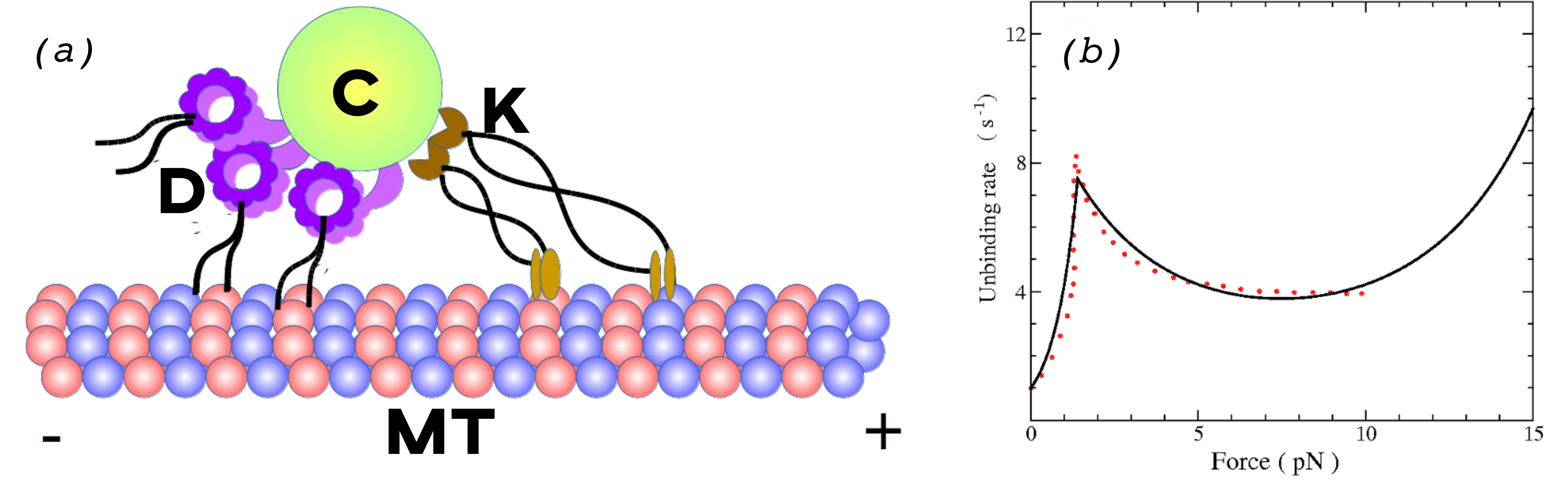}
    \caption{(a) Schematic of bidirectional motion of cargo (C) attached to both kinesin (K) and dynein (D) motors on a microtubule (MT) filament; (b) Single dynein unbinding rate from experiments \cite{kunwar2011mechanical} (points) and the corresponding fit (solid line) from the TFBD model \cite{nair2016effect}. }
    \label{fig:Fig1}
\end{figure*}

Bidirectional transport is ubiquitous in nature in the context of intracellular transport \cite{welte2004,gross2002interactions,welte1998developmental,hollenbeck1996pattern}. Within the cell, oppositely directed motor proteins such as dynein and kinesin motors walk on microtubule (MT) filaments \cite{welte2004,schuster2011transient} to transport diverse cellular cargo \cite{welte2004}. A theoretical framework proposed to explain the bidirectional transport is based on the {\it tug-of war} hypothesis \cite{welte1998developmental,schuster2011transient,welte2004,hancock2014bidirectional,muller2008tug,muller2008motility,muller2010bidirectional,Bhat2012}, which posits that the motors stochastically binds to and unbinds from the filament while mechanically interacting with each other  through the cargo that they carry (Fig.~\ref{fig:Fig1}a) \cite{hancock2014bidirectional,muller2008tug,muller2008motility}. The resultant motion arises due to the competition between the oppositely directed motors \cite{muller2008tug,muller2008motility}.

The  {\it tug-of-war} model predicts that inhibiting the activity of one motor species would lead to an enhancement of motility in the other direction. While many experiments have provided support for this mechanical tug-of-war picture \cite{schuster2011transient,muller2008tug,hendricks2010motor,gennerich2006finite,soppina2009tug}, there remain a large class of experiments which are incompatible with the predictions of this model and show that there exists some coordination mechanism due to which inhibition of one motor species results in an overall decline in the motility of the cargo \cite{hancock2014bidirectional,gross2002interactions,martin1999cytoplasmic,ally2009opposite,gross2002coordination,EncaladaCell2011}. This apparently counterintuitive finding has been referred to as the {\it paradox of codependence} \cite{hancock2014bidirectional,welte2004}. The resolution of this paradox in  terms of the underlying mechanisms which govern bidirectional transport remains an important open question.

Unlike kinesin, whose detachment rates from the filament {\it increases} exponentially with increasing load force - a characteristic of {\it slip bond} \cite{pereverzev2006force,leidel2012measuring,klumpp2005cooperative}, dynein motors exhibit {\it catchbonding} : the propensity for the dynein motors to unbind {\em decreases} when subjected to increasing load forces in certain force regimes (Fig1b) \cite{kunwar2011mechanical,mallik2013teamwork,leidel2012measuring}.

While the effect of catchbonding  has previously been incorporated  in context of  modeling of bidirectional transport of lipid droplets \cite{kunwar2011mechanical}, their importance in mediating codependent transport properties has not been realized and investigated. In this article we study the generic mechanism by which catchbonding in dynein may manifest as codependent transport behaviour for cellular cargoes and quantify the effects of the catchbond in terms of experimentally measurable cargo transport characteristics. In particular we explicitly show that catchbonding in dynein provides one plausible means of resolving the {\it paradox of codependence}.

 We use a threshold force bond deformation (TFBD) model to fit the experimentally observed unbinding rate of single dynein motors (Fig.~\ref{fig:Fig1}(b)) \cite{nair2016effect}. With the TFBD model for dynein, and the usual slip bond model for kinesin \cite{klumpp2005cooperative,muller2008tug}, we study the transport properties of bidirectional cargo motion by multiple motors, using experimentally relevant measures : (i) average processivity, defined as the mean distance a cargo travels along a filament before detaching, (ii) probability distributions of runtime and pause times, and (iii) typical cargo trajectories as well as distributions of cargo velocities. Using these measures we show that, in an experimentally viable parameter space, the catchbonded response of dynein provides an internal regulatory mechanism that exhibits codependent transport characteristics.

\section{Theory and Simulation}

\subsection{Model}

    We consider transport of a cellular cargo with $N_+$ kinesin motors and $N_-$ dynein motors. These motors stochastically bind to a MT filament with rates $\pi_{\pm}$ and unbind with rates $\varepsilon_{\pm}$. 
At any instant of time, the state of the cargo is characterized by the number of attached Kinesin ($n_+$) and Dynein motors ($n_-$). The maximum of number of kinesin and dynein motors are $N_+$ and $N_-$ respectively ($0 < n_+ < N_+$ and $0 < n_- < N_-$). The time evolution of the system is then governed by the master equation \cite{muller2008tug}

	\begin{widetext}
	\begin{eqnarray}
	\frac{\partial p(n_+,n_-)}{\partial t} &=& p (n_+ + 1, n_-) \epsilon_+(n_+ + 1, n_-) + p (n_+, n_- + 1) \epsilon_-(n_+, n_- + 1) \nonumber \\
	&&+ p(n_+ - 1, n_-) \pi_+(n_+ - 1, n_-) + p(n_+, n_- - 1) \pi_-(n_+, n_- - 1) \nonumber \\
	&& - p(n_+,n_-) \left[   \epsilon_+(n_+, n_-) + \epsilon_-(n_+, n_-) + \pi_+(n_+, n_-) + \pi_-(n_+, n_-) \right]
	\end{eqnarray}	
	\end{widetext}	
	where, $p(n_+,n_-)$ is the probability to find the cargo with $n_+$ kinesin and $n_-$ dynein motors.

The kinesin and dynein binding rates are assumed to be of the form $\pi_{\pm} = (N_{\pm} - n_{\pm}) \pi_{0\pm}$, where $N_+ \pi_{0+}$ ($N_- \pi_{0-}$) is the rate for the first kinesin (dynein) motor to bind to the MT. 

 Dynein motors exhibit catchbonding at forces larger than the stall force, $F_{s-}$, defined as the load force at which the cargo stalls \cite{kunwar2011mechanical,mallik2013teamwork,leidel2012measuring}.  This catchbonding regime is characterized by a decreasing detachment rate with increasing opposing load (see Fig.~\ref{fig:Fig1}(b)). The load force is assumed to be shared equally among the attached motors. We use the phenomenological TFBD model for the unbinding rate of a dynein in an $(n_+,n_-)$ state  \cite{nair2016effect, Chakrabarti2017}, given by
    \begin{equation}
    \varepsilon_- = n_- \varepsilon_{0-} \exp [-E_d(F_c) + F_c/(n_- F_{d-})]
    \end{equation}
    where the deformation energy $E_d$ sets in at $F > F_{s-}$, and is modeled by a phenomenological equation \cite{nair2016effect},
    \begin{equation}
    E_d(F_c) = \Theta (F_c/n_- - F_{s-})  \alpha \left[1 - \exp\left(-\frac{F_c/n_- - F_{s-}}{F_0}\right)\right]
    \end{equation}
    The parameter $\alpha$ sets the strength of the catch bond, while $F_{d-}$ and $F_0$ characterize the force scales for the dissociation energy and the deformation energy respectively, while  $F_c$ is the cooperative force felt by the motors due to the effect of the motors of the other species.  Unlike dynein, the unbinding kinetics of kinesin exhibits usual {\em slip} behavior, and thus the unbinding rate for kinesin is given by the expression $\varepsilon_+(n_+,n_-) = n_+ \varepsilon_{0+} \exp [F_c(n_+,n_-)/(n_+ F_{d+})] $ \cite{muller2008tug}. The characteristic stall forces and detachment forces of kinesin are denoted by $F_{s+}$ and $F_{d+}$ respectively. 
	
 The expression for the cooperative force felt by the motors is given by \cite{muller2008motility}
	\begin{equation}
	F_c(n_+,n_-) = \frac{n_+ n_- F_{s+} F_{s-}}{n_- F_{s-} v_{0+} + n_+ F_{s+} v_{0-}} \left( v_{0+} + v_{0-} \right)
	\end{equation}
	and the cargo velocity is given by
	\begin{equation}
	v_c(n_+,n_-) = \frac{n_+ F_{s+} - n_- F_{s-}}{n_- F_{s-} / v_{0-} + n_+ F_{s+} / v_{0+}}
	\end{equation}
	Here, $v_{0\pm}$ denotes the velocity of kinesin (or dynein) motors,
	\begin{equation*}
	v_{0+} = \left \lbrace \begin{array}{ccc}
	v_{F+} & if  & v_c > 0 \\
	v_{B+} & if & v_c < 0
	\end{array}\right .
	\mathrm{and}\;
	v_{0-} = \left \lbrace \begin{array}{ccc}
	v_{F-} & if & v_c < 0 \\
	v_{B-} & if & v_c > 0
	\end{array}\right .
	\end{equation*}
	where, $v_F$ and $v_B$ are the forward and backward motor velocities. Finally the stall forces for the two motor species are denoted by $F_{s\pm}$.


    \begin{table}[t!]
		\begin{tabular}{||c|c|c|c|c|c||}
			\hline
			\hline
			Parameter & Kinesin & Ref. & Dynein & Ref. \\
			\hline
			$F_{s\pm}$ & 6 pN & \cite{schnitzer2000force} & 1 pN (Weak) & \cite{mallik2005building} \\
			
			& & & 7 pN (Strong) & \cite{toba2006overlapping} \\
			\hline
			$F_{d\pm}$ & 3 pN & \cite{schnitzer2000force} & 0.67 pN & \cite{kunwar2011mechanical,nair2016effect} \\
			\hline
			$\pi_{0\pm}$ & 5/s & \cite{beeg2008transport} & 1/s & \cite{leduc2004cooperative} \\
			\hline
			$\varepsilon_{0\pm}$ & 1/s & \cite{schnitzer2000force} & (0.1 - 10)/s & \cite{reck2006single} \\
			\hline
			$v_{F\pm}$ & $0.65 \mu m/s$ & \cite{carter2005mechanics} & $0.65 \mu m/s$ & \cite{king2000dynactin} \\
			\hline
			$v_{B\pm}$ & $1 nm/s$ & \cite{carter2005mechanics} & $1 nm/s$ & \cite{kojima2002mechanical,gennerich2007force} \\
			\hline
			\hline
		\end{tabular}

		\caption{Single motor parameter values used in the simulations.}
		\label{tab1}
	\end{table}
	The parameters used in the study are taken from the literature, and are summarized in Table~\ref{tab1}.

\subsection{ First Passage Time and Processivity}
	
	The Mean First Passage time (MFPT) in a particular bound motor state $(n_+,n_-)$, $T_{n_+,n_-}$, is defined as the mean time for cargo starting with $n$ bound kinesins and $m$ bound dyneins to unbind, i.e; reach the $(0,0)$ state. This can be expressed in terms of mean residence time in that state $\tau_{n_+,n_-}$ and transition probabilities to other states, which leads to a recursion relation for the MFPT, of the form 
	\begin{eqnarray}
		T_{n_+,n_-} &=&  \tau_{n_+,n_-} \left( 1  +  \pi^{+}_{n_+,n_-}T_{n_+ +1,n_-}  + \pi^{-}_{n_+,n_-}T_{n_+,n_-+1} \right.\nonumber\\ & +& \left. \varepsilon^{+}_{n_+,n_-}T_{n_+-1,n_-}  + \varepsilon^{-}_{n_+,n_-}T_{n_+,n_--1} \right) 
	\end{eqnarray}
	where the mean residence time in a $(n_+,n_-)$ state is simply the inverse of the sum of the transition probabilities to the other states, $\tau_{n_+,n_-} = 1/( \pi^{+}_{n_+,n_-} +  \pi^{-}_{n_+,n_-} + \varepsilon^{+}_{n_+,n_-} +  \varepsilon^{-}_{n_+,n_-})$.
	
We can similarly develop a recursion relation for the average cargo processivity (ACP) $L_{n_+,n_-}$, defined as the average distance a motor starting from the (n,m) state walks before it unbinds. In the state $(n_+,n_-)$, the cargo walks with the cooperative velocity, $ v_{c}(n_+,n_-)$, and the mean residence time in this state is $\tau(n_+,n_-)$. Hence the mean distance $\eta_{n_+,n_-}$ that the cargo walks in the $(n_+,n_-)$ state before transition to another state can be expressed as $\eta_{n_+,n_-} = v_{c}(n_+,n_-) \tau_{n_+,n_-}$. With this identification, the recursion relation for the mean processivity becomes,

	\begin{eqnarray}
	L_{n_+,n_-} &=& \eta_{n_+,n_-}\left(1 + \pi^{+}_{n_+,n_-}L_{n_++1,n_-} + \pi^{-}_{n_+,n_-}L_{n_+,n_-+1}\right. \nonumber \\
	&+& \left. \varepsilon^{+}_{n_+,n_-}L_{n-1,m} + \varepsilon^{-}_{n_+,n_-}L_{n_+,n_--1}\right) 
	\label{eq:mcp}
	\end{eqnarray}
	 Together with the absorbing boundary conditions, $T_{0,0}=0$ and $L_{0,0}=0$, these define a linear system of equations which can be solved analytically to obtain the MFPT and the ACP.
	 
	 The average processivity reported in this manuscript is the average over all possible initial states of the motor conformations for a given maximum number of kinesins and dyneins, 
	 \begin{equation}
		 \langle L_{n_+,n_-} \rangle_{n_+n_-} = C \sum_{n_+ = 0}^{N_+} \sum_{n_- = 0}^{N_-} L_{n_+,n_-} \left( 1 - \delta_{n_++n_-,0}\right),
\label{eq:Lavg}
	 \end{equation}
where $C$ is a normalization factor which depends on $N_+$ and $N_-$, with $ C^{-1} = (N_+ + 1)(N_- +1) -1$.

\subsection{Simulations and Numerical techniques}

The Master equation is simulated using Stochastic Simulation Algorithm (SSA) \cite{gillespie1976general,gillespie1977exact} to obtain individual cargo trajectories. All possible initial configurations were generated for a $(N_+, N_-)$ pair, and 1000 trajectories were evolved for each initial configuration. A run finishes if the simulation continues until the maximum time $T_{MAX} \sim 10^{4} s$ or if all motors detach from the MT. The runlength was then averaged over all initial configurations and all iterations. Probability distributions were also computed from the SSA trajectories after discarding initial transients. The simulated trajectories are then analysed to quantify the statistical properties of the system. Further we perform Brownian dynamics simulations and determine processivity of the cargo when the load is shared stochastically (see Appendix A).

Further we also  derive the associated Fokker-Planck equation (FPE) corresponding to the underlying Master equation, by treating the number of attached motors as continuum variables in the state space (see Appendix B).	




    \begin{figure*}[t]
    	\centering
    	\includegraphics[width=0.9\linewidth]{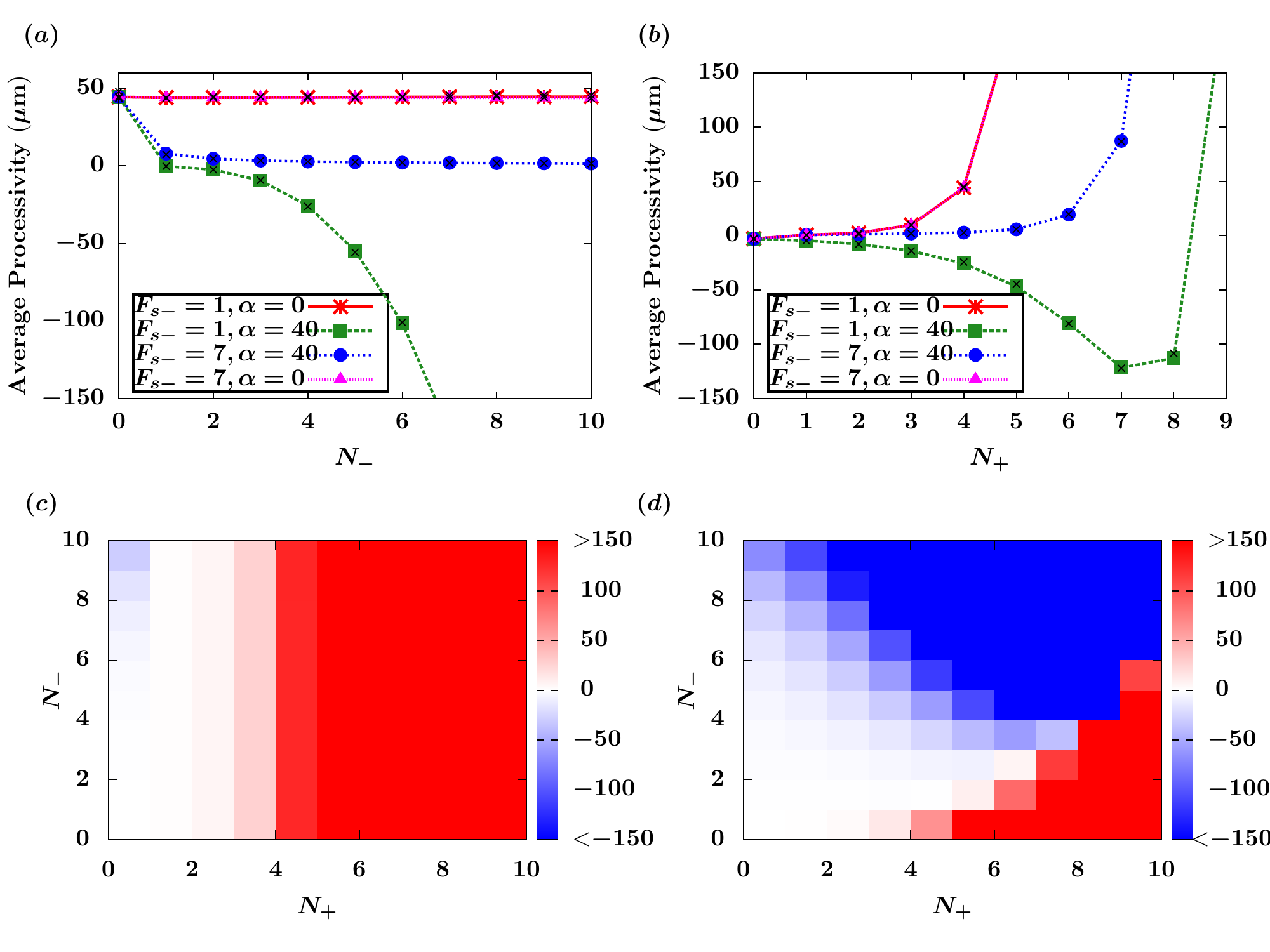}
    	\caption{Average processivity (a) as a function of $N_-$ for $N_+ = 4$ , and (b) as a function of $N_+$ for $N_- = 4$. The colored points and lines correspond to the simulation results. Black crosses in all cases are obtained by the solutions of Eq.~\ref{eq:Lavg}. Contour plots for processivity obtained from Eq.~\ref{eq:Lavg} in the $N_+ - N_-$ plane for (c) $F_{s-} = 1 pN, \alpha = 0,  F_0 = 7pN$, and (d) $F_{s-} = 1 pN, \alpha = 40 k_B T, F_0 = 7pN$. The color bar indicates the average processivity (in $\mu m$). 
	    The zero-force (un)binding rates for dynein are $\varepsilon_{0-} = \pi_{0-} = 1/s$}
    	\label{fig:npnm}
    \end{figure*}

\section{Results}

\subsection {Cargo Processivity Characteristics} 

We show results for the average processivity $\langle L_{n,m} \rangle_{nm}$ with varying $N_{-}$ (Fig. ~\ref{fig:npnm} (a)) and $N_{+}$ (Fig. ~\ref{fig:npnm} (b))  for “weak” dynein (mammalian, $F_{s-}=1pN$) and “strong” dynein (yeast, $F_{s-}=7pN$). For all cases, the analytical value of the average processivity shows excellent agreement with SSA results. In Fig.~\ref{fig:npnm} (a), we observe a sharp decrease in plus-end directed processivity with increase in $N_{-}$, due to an increased propensity of catchbonded dynein motors to latch on to the filament. For weak dynein, the cargo in fact  reverses direction, due to the activation of catchbond at lower forces.

\subsection { Resolution of Paradox of Codependence}

 \begin{figure*}[!t]
 	\centering
 	\includegraphics[width=0.9\linewidth]{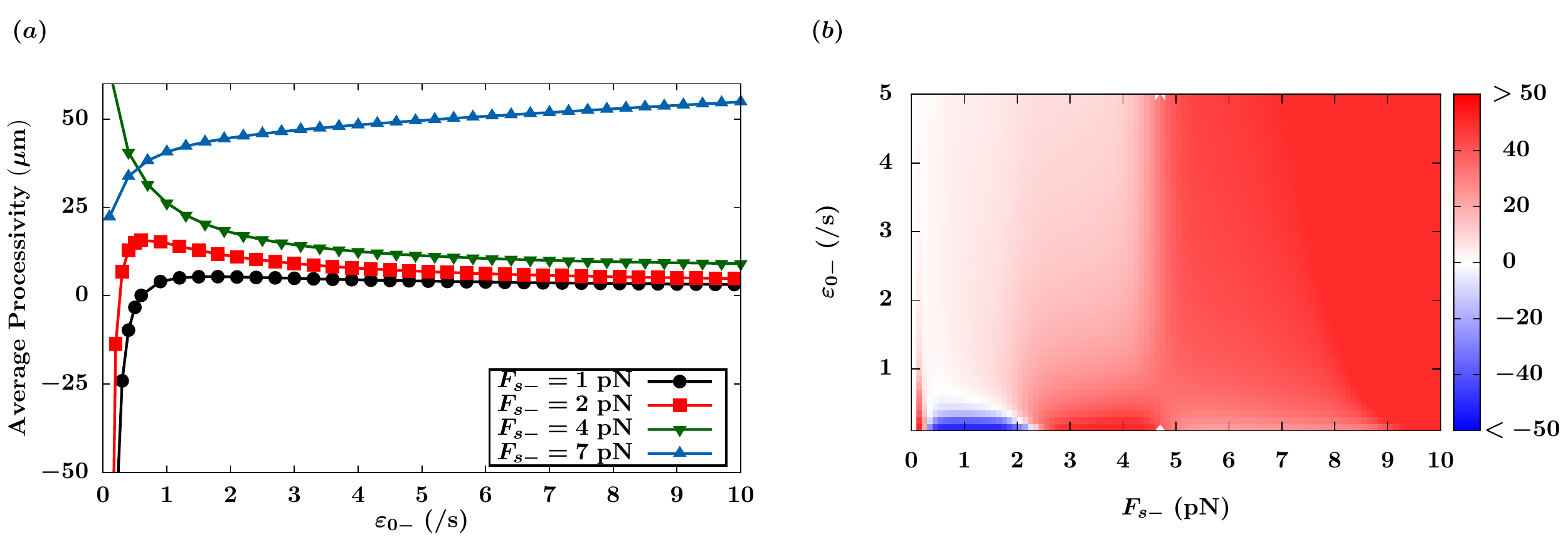}
 	\caption{(a) Average processivity as a function of $\varepsilon_{0-}$ for different stall forces at $\alpha = 40 k_B T$; obtained using Eq.~\ref{eq:Lavg}. (b) Contour plots of processivity in $(F_{s-}-\varepsilon_{0-})$ plane for $\alpha = 40 k_B T$ and $F_0 = 7pN$. Data shown is for $N_+ = 6$, $N_- = 2$,  $\pi_{0-} = 1/s$.}
 	\label{fig:fsmepsm}
 \end{figure*}
    
 Diverse experiments have indicated that mutations of conventional kinesin in {\it Drosophila} can hamper motion of cellular cargo in both directions, by effectively reducing the number of motors attached to the cargo \cite{saxton1991kinesin,hurd1996kinesin,gindhart1998kinesin,ally2009opposite,gross2002coordination}.
While the conventional tug-of-war model without the incorporation of catchbond does not exhibit codependent transport characteristics and fails to resolve the {\it paradox of codependence} observed in these experiments, the processivity characteristics  reveals clear signature of plausible resolution of this Paradox by means catchbond mediated mechanism. To investigate this, in Fig.~\ref{fig:npnm} (b), we look at the effect of variation of $N_{+}$ on processivity, for a fixed value of $N_-$. Remarkably, the average processivity for weak dynein shows a non-monotonic behaviour with increasing $N_+$. In particular {\it there is decrease of processivity in the negative direction on decreasing the number of plus-end directed motors.} This  is a singular feature arising solely due to catchbonding in dynein, contrary to usual tug-of-war predictions, and is reminiscent of the {\em paradox of codependence}. The robustness of this catchbond mediated phenomenon can be gauged from the observation that it persists for a wide range of biologically relevant parameters even when the load is shared stochastically between the motors (see Appendix A and Fig.~8 for details).

This codependent behaviour exemplified in processivity characteristics may be understood in terms of the catchbond mechanism at play. In the absence of opposing load, increasing $N_+$ has the effect of increasing the mean first passage time (MFPT) for the kinesin motors. However in the presence of dynein, with larger number of kinesins, the load per dynein is higher, leading to engagement of the catchbond and thus fewer detachment events for dynein. The cargo is now in a tug-of-war state, leading to higher detachment forces on the opposing kinesins, which detach with the usual slip kinetics. Thus, on average, for some parameter regime, the kinesins detach at a higher rate than dyneins, leading to more configurations where there are no kinesins opposing the dynein team. Thus although the direct effect of the catch bond is a larger value of average unbinding time for dyneins, this leads to more configurations where the dyneins can walk towards the negative end leading to codependent transport.

 The corresponding contour plots of the processivity of the cargo, which provide an experimental testbed, in the $(N_+ - N_{-})$ plane are shown in Fig.~\ref{fig:npnm}(c-d), for ‘weak’ dynein where the effect of dynein catch-bond is robust. As expected, in the absence of catch-bond ($\alpha = 0$) (Fig.~\ref{fig:npnm}(c)), there is a smooth transition from negative-directed runs to positive directed runs. In the presence of catch-bonded dynein (Fig.~\ref{fig:npnm}(d)), we observe a distinct regime where the processivity increases in the negative direction on increasing $N_+$, reminiscent of anomalous codependent transport. Plus-end directed motion now occurs only for large $N_+$ and low $N_-$. This non-trivial effect of the catch bond is a robust feature that is observed for other values of kinesin and dynein motors (see Appendix C, Figs. 9 and Fig. 10) and can also be understood in terms of the average number of bound motors (see Appendix D, Fig. 11).
 
Experimental techniques to modulate cargo processivity can also be achieved by modifying the (un)binding rates of the motor proteins. Dynactin mutations in {\em Drosophila} neurons affect the kinetics of dynein binding to the filament, leading to cargo stalls \cite{martin1999cytoplasmic}.  Similarly, the tau protein has been observed to change the unbinding rates of kinesin and dynein motors \cite{hendricks2017tau}. To investigate this, we look at the effect of variation of the bare unbinding rate of dynein motor on processivity of the cargo ($\varepsilon_{0-}$) (Fig.~\ref{fig:fsmepsm}a). Codependent transport behaviour is again observed for a range of stall forces. For instance  at $F_{s-} = 2pN$, we observe a non-monotonic behaviour of the processivity with increasing unbinding rate. At $F_{s-} = 4pN$, the run length in the positive direction decreases on increase in $\varepsilon_{0-}$.  The contour plot of the processivity in the $(F_{s-}-\varepsilon_{0-})$ plane (Fig.~\ref{fig:fsmepsm}(b))  shows non-monotonic signatures of codependent transport - a feature akin to reentrant behaviour \cite{narayanan1994reentrant} -  for a range of stall forces, and highlights the role of dynein stall force in determining the overall motion of the cellular cargo. The strength of catchbond ($\alpha$)  plays an important role in determining the {\it nature} of processivity of the cargo (Appendix E, Fig. 12). A microscopic modeling of the catch bond in dynein based on the experimentally determined mechanism of the catch bond \cite{gennerich2015PNAS} can help identify biologically relevant regimes for $\alpha$ and $F_o$ and therefore constrain the predictions of the model.

\subsection{Probability distribution of runtime and cargo velocities}
    \begin{figure*}[t]
	    	\centering
	    	\includegraphics[width=\linewidth]{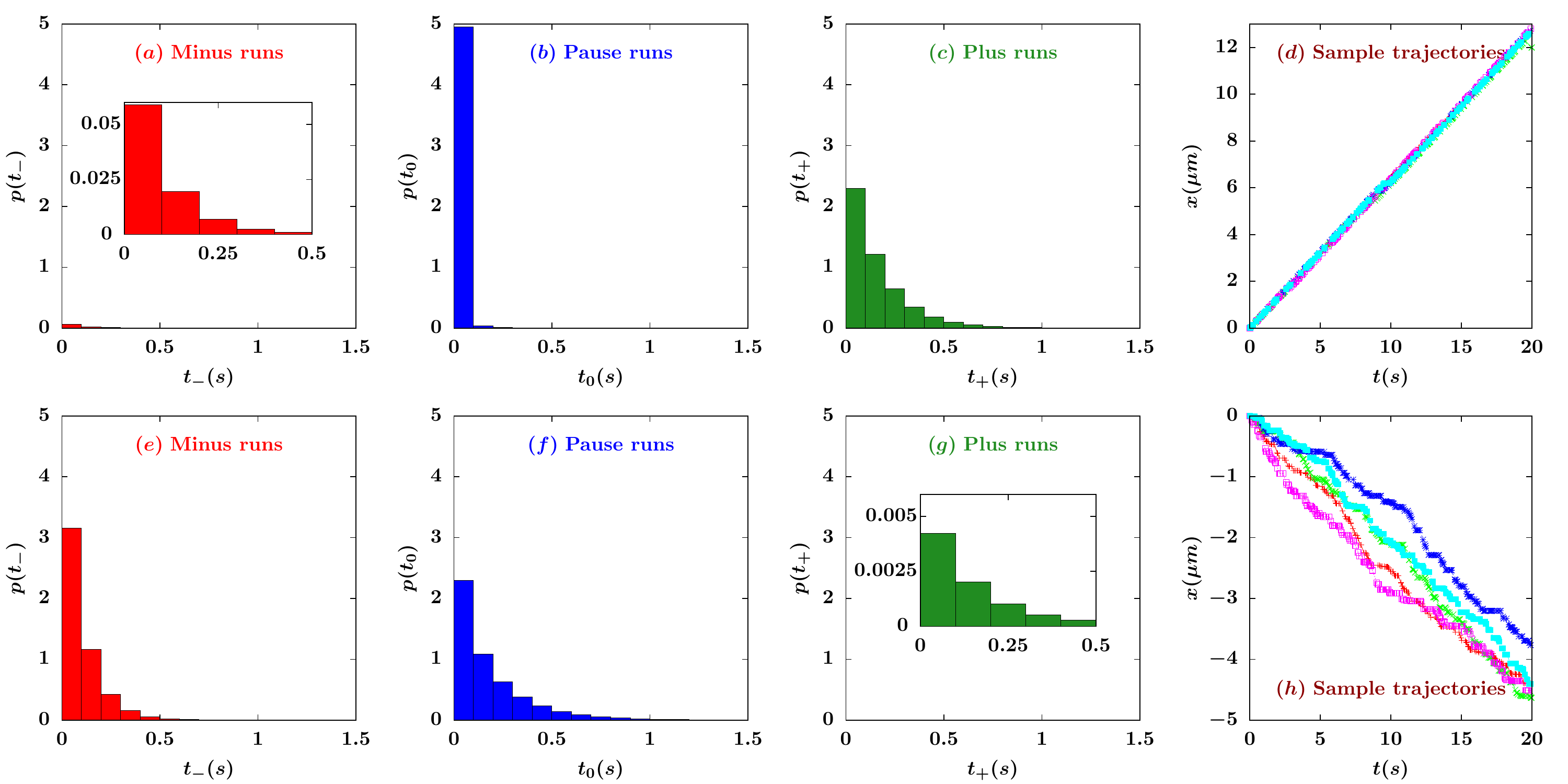}
	    	\caption{Probability distributions of runtime for $N_+ = 2$, $N_- = 6$. The top panels show the normalized histograms and sample trajectories for dynein in the absence of catch bond ($\alpha = 0$). The bottom panels show the corresponding quantities for catch-bonded dynein ($\alpha = 40$, $F_0 = 7pN$). (a) and (e) Distributions of runtime for minus directed runs (shown in red); (b) and (f) pausetime distributions (shown in blue)(c) and (g) distributions of runtime for plus directed runs (shown in green); and  (d) and (h) sample trajectories. Insets in (a) and (g) show magnified views of the corresponding distributions.}
	    	\label{fig:hist26}
	 \end{figure*}

 \begin{figure}[t]
	    	\centering
	    	\includegraphics[width=\linewidth]{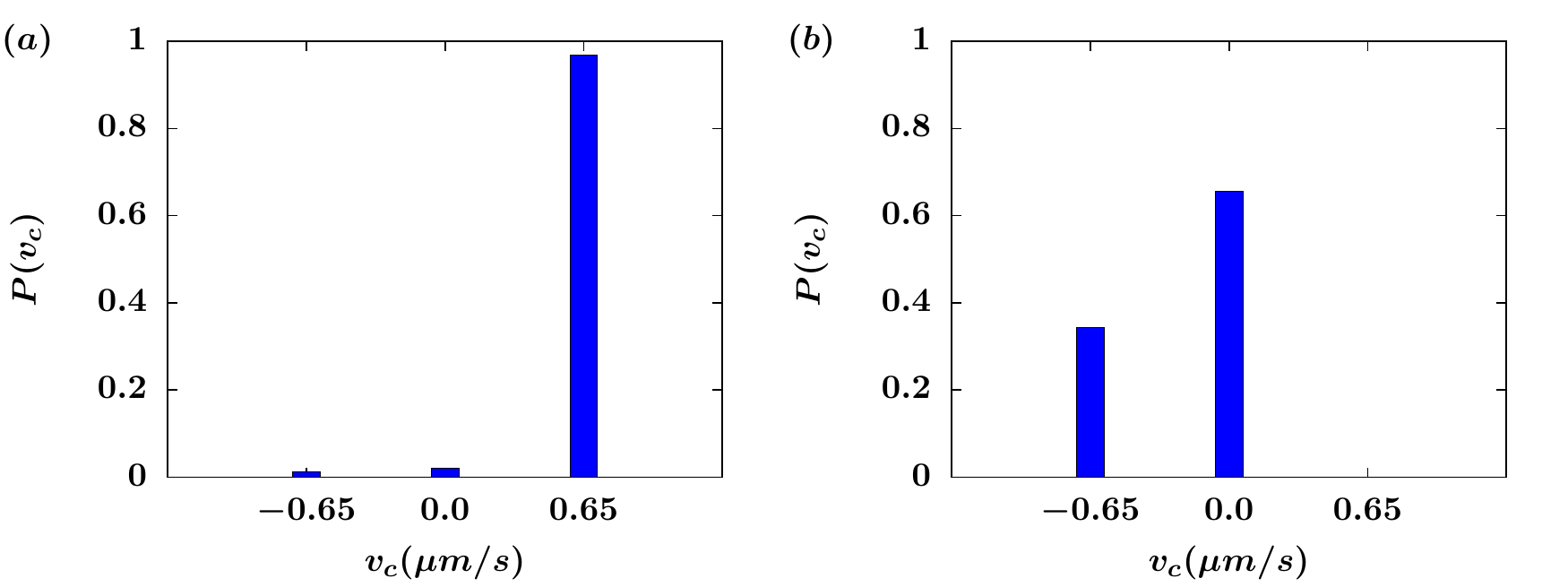}
	    	\caption{Probability distributions of velocities for $N_+ = 2$, $N_- = 6$.The left panels show the normalized histograms in the absence of catch bond ($\alpha = 0$). The right panels show the corresponding histograms when dynein is catch-bonded ($\alpha = 40$, $F_0 = 7pN$).}
	    	\label{fig:hist26v}
	 \end{figure}

\begin{figure*}[t]
   	\centering
   	\includegraphics[width=\linewidth]{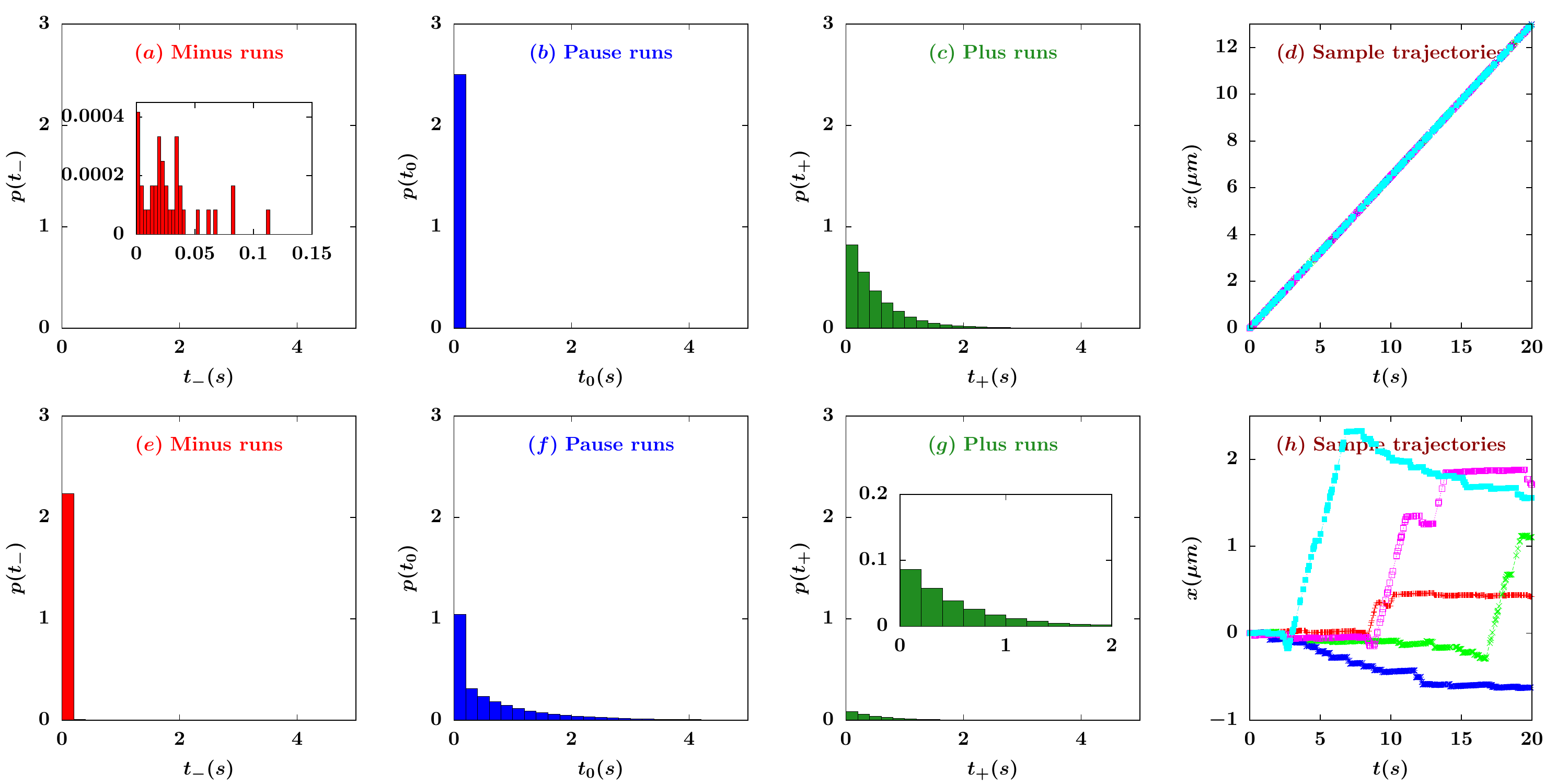}
   	\caption{Probability distributions of runtime for $N_+ = 6$, $N_- = 2$. The top panels show the normalized histograms and sample trajectories for dynein in the absence of catch bond ($\alpha = 0$). The bottom panels show the corresponding quantities for catch-bonded dynein ($\alpha = 40$, $F_0 = 7pN$). (a) and (e) Distributions of runtime for minus directed runs (shown in red); (b) and (f) pausetime distributions (shown in blue); (c) and (g) distributions of runtime for plus directed runs (shown in green); and  (d) and (h) sample trajectories. Insets, where present, show a magnified view of the probability distributions.}
   	\label{fig:hist62}
 \end{figure*}

 In order to highlight the role of catchbond we provide quantitative measures which are biologically relevant for comparison with experimental data related to trajectories of cellular cargo carried by molecular motors. We analyze the probability distribution of the time the cargo spends in the paused ({\it tug-of-war}) state versus the time it spends in the moving plus-end directed and minus-end directed state, as well as the probability distribution of the velocities of the cargo.

Motivated by experiments on {\it dictyostelium} cell extracts \cite{soppina2009tug}, we study the transport behaviour of a cargo with $N_+ = 2$ and $N_- = 6$ (Fig.~\ref{fig:hist26} and Fig.~\ref{fig:hist26v}).  In the absence of catchbonding, cargoes predominantly move with positive velocity and the resultant motion is strongly plus-end directed (Fig.~\ref{fig:hist26}d). The probability distributions of runtime show that there are many more kinesin runs (Fig.~\ref{fig:hist26}c) than dynein runs (Fig.~\ref{fig:hist26}a), and the average runtime is also higher in the case of kinesins. The pauses in this case are also of extremely short duration (Fig.~\ref{fig:hist26}b). The corresponding probability distribution for the velocities are shown in Fig.~\ref{fig:hist26v}a

In contrast, when dynein catch bond is switched on, the picture changes dramatically. While the cargo is in a paused state a significant fraction of time, around $35\%$ of its runs are negative directed (Fig.~\ref{fig:hist26}f inset). Minus-ended runs become much more frequent than plus-ended runs, and the cargoes tend to move with a negative velocity (Fig.~\ref{fig:hist26v}b) while the average pause time also increases by an order of magnitude compared to the non-catchbonded case, and becomes comparable to the average minus directed runtime. This is shown in Figs.~\ref{fig:hist26}(e)-(g). This prediction of minus-ended runs with intermittent pauses qualitatively agrees with the  experimental observation of transport of endosomes in {\it Dictyostelium} cells \cite{soppina2009tug}. 
   
   In a separate set of experiments on early endosomes in fungi, a team many kinesin motors (3-10)  are involved in {\it tug-of-war} with  1 or 2 dynein motors during transport \cite{schuster2011transient}. The results displayed  in Fig.~\ref{fig:hist62} for a cargo being transported by six kinesins and two dyneins illustrates that while in the absence of catchbonding in dynein, the resultant motion would be strongly plus-end directed, with very small pause times, incorporation of catchbonding results in the frequency of minus-ended runs exceeding the frequency of plus-ended runs by almost one order of magnitude. However, the average duration of the minus-ended runs is about one order of magnitude lower than that of the plus-end directed run duration. Further there are now substantial duration of pauses ($1-4$ sec) during transport. These characteristics of the probability distributions result in typical cargo trajectories which exhibits bidirectional motion with pauses. The role of dynein catchbonding in altering the transport characteristics can also be seen  for the simplest possible case of bidirectional transport of a cargo by a single kinesin and a single dynein motor (Appendix F, Fig. 13).

\subsection{Quantitative comparison with experiments}

 \begin{figure*}[!t]
 	\centering
 	\includegraphics[width=0.8\linewidth]{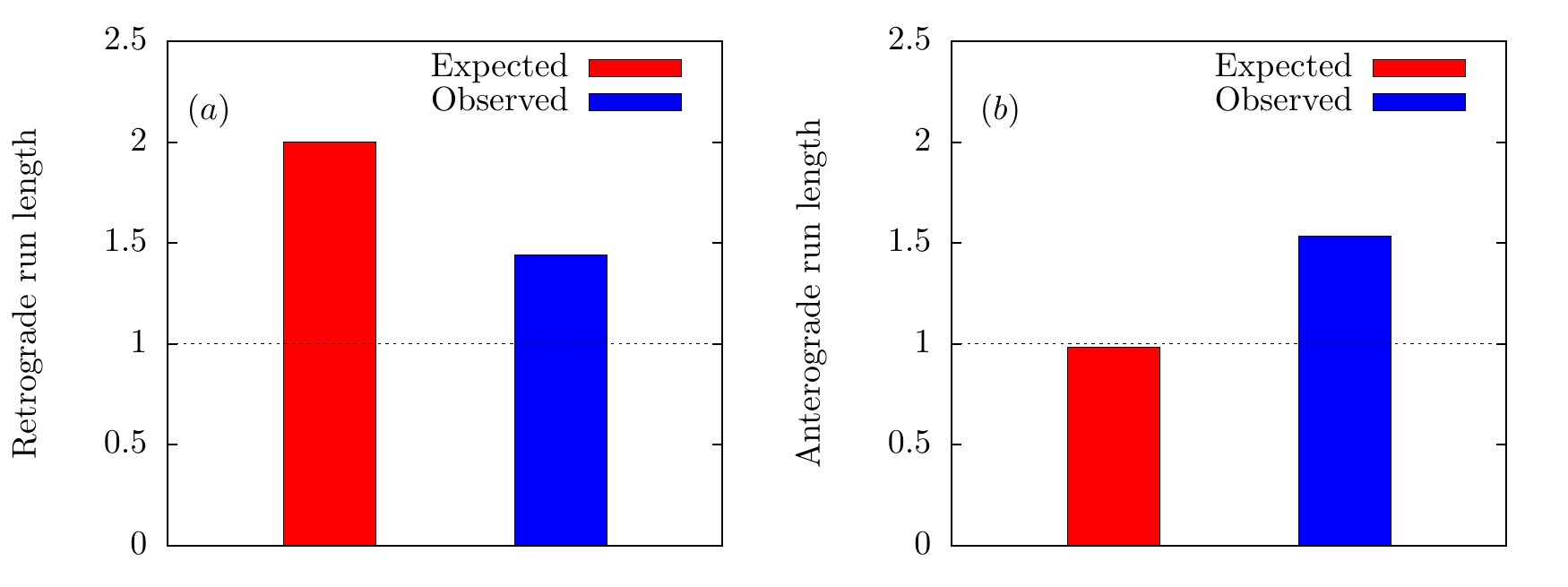}
	 \caption{ Histograms showing  scaled (a) retrograde and (b) anterograde run lengths with {\it non-catchbonded} (Expected, red, $\alpha = 0$) and {\it catchbonded} (Observed, blue, $\alpha = 40 k_B T$, $F_0 = 7pN$ ), when $N_+$ is changed from $3$ to $2$, while $N_- = 4$. The scaling is done with respect to the control ( without kinesin inhibition) and corresponds to $N_+ = 3$. The zero-force (un)binding rates for dynein are $\varepsilon_{0-} = \pi_{0-} = 1/s$ }
 	\label{fig:comp}
 \end{figure*}

In order to provide a quantitative comparison of our results with in-vivo experiments, we consider the specific case of kinesin inhibition in mouse neurons~\cite{EncaladaCell2011}. It was observed that inhibiting kinesin resulted in smaller retrograde run lengths of prion protein vesicles, which is contrary to expectations - a signature of codependent transport behaviour. In our model, kinesin inhibition is incorporated by reducing the number of kinesins $(N_+)$ from $3$ to $2$ while the dynein number is held fixed $(N_- = 4)$. As shown in Fig.~\ref{fig:comp}, this reduction in kinesin motors leads to smaller retrograde run lengths and larger anterograde run lengths when catch bond is switched on in dynein, as opposed to the situation when dynein unbinding exhibits slip behavior. This is the scenario of co-dependent transport and compares well with the experimental observations. Our assumption that kinesin inhibition leads to reduction in its number is a simplified view of the effect of the inhibition experiment in in-vivo conditions. Nonetheless, even with this assumption our results definitively points to the role of catchbond mediated mechanism in determining codependent transport behaviour.

\section{Conclusion}
   In summary, the findings of our model points to the crucial role played by catchbonding in dynein motors in internally regulating transport and providing a possible resolution of the paradox of codependence. It also provides a framework to interpret diverse set of experiments where regulation of transport is achieved by different modes of modification of the motor properties. 
   For instance, while decreasing $N_-$ or increasing $\varepsilon_{0-}$ has the effect of weakening the dynein motor action, the manifestation of these two effects in the transport characteristics can in general be distinct. The results of these experiments can then qualitatively be understood in the light of Fig.~\ref{fig:npnm} and Fig.~\ref{fig:fsmepsm}, where weakening the dynein motor can lead to stalled motion of the cargo. 
   Interestingly, while kinesin exhibits a conventional slip bond, the cooperative force exerted by the catch bonded dynein on kinesins, and vice-versa, introduces a complex interplay which results in signatures of codependent transport being observed even on varying effective kinesin numbers.
This effect is reflected in a preliminary comparison of processivity measurements for  prion protein vesicles in mouse neurons \cite{EncaladaCell2011} with our model predictions.

   
 These processivity measures also point to the sharp difference in transport characteristics for strong and weak dynein. In the former case, regulatory role of catchbonding is very weak due to the high force scale at which catchbond is activated. This may provide a clue as to why the {\it strong} dynein in yeast is not involved in transport, while {\it weak} mammalian dynein are crucial to intracellular transport.
   
   Apart from the internal regulatory mechanism described here, external regulation by associated proteins is also expected to play an important role in determining the transport characteristics. Various candidate proteins such as {\it Klar} and JIP1 have been shown to modify transport behaviour \cite{welte2004,gross2003dynactin,gross2004hither,gross2003determinant,welte1998developmental,gross2000dynein,gross2002coordination,shubeita2008consequences,welte2005regulation,mckenney2010lis1,fuholzbaurJCB2013}. Further, various other factors, such as memory effects during motor rebinding \cite{leidel2012measuring}, interactions between multiple motors \cite{lipowskyelastic1, lipowskyelastic2}, variable dynein step sizes \cite{mallik2004cytoplasmic,kunwar2011mechanical}, and stochastic load sharing could also modify the transport behaviour of the cargo. However we show using simulations incorporating a stochastic sharing of load between attached motors, that the codependent behavior of cargo processivity is robust and is preserved even with additional inputs such as viscous friction and thermal noise (see Supplementary section I(E)).
   
   Various regulatory mechanisms are expected to achieve coordination through different means which may be reflected in the transport characteristics of the cargo. For example, in the case of the catch-bonded tug-of-war mechanical model, the pause state would in general be characterized by a slow velocity of the cargo. On the other hand, for mechanical inhibition \cite{mechanical, hancock2014bidirectional}, microtubule tethering mechanism \cite{tethering, hancock2014bidirectional} or steric disinhibition\cite{steric, hancock2014bidirectional}, the motion of the cargo would either be diffusive or would show no movement. Increasing the binding rates of either motor species would result in shorter pause times if coordination is achieved through mediation by the catch bond, while it would have no effect on the pause times for some other mechanism. A careful examination of high resolution spatio-temporal measurement of cargo processivity and pause durations obtained in various experiments is required to delineate the relative importance of these internal regulatory mechanisms.
   
   To conclude, we show that catchbonding in dynein dramatically alters the transport characteristics, and manifests as an internal regulatory mechanism that provides one possible resolution of the {\it paradox of codependence}.  
   
   
\vskip 0.5cm
\noindent
{\em Acknowledgements.} Financial support is acknowledged by MKM for Ramanujan Fellowship (13DST052), DST and  IITB (14IRCCSG009); AC for SERB project No. EMR/2014/000791 and DST; SM and MKM for SERB project No. EMR /2017/001335; NG for UGC and SC for DST. MKM acknowledges hospitality of MPIPKS, Dresden. SM and MKM acknowledge helpful discussions with Roop Mallik, TIFR. 

\vskip 0.5cm
\noindent
P.P and N.G carried out simulations and analyzed data. S.C, S.N., and A.N. carried out simulations. A.C., M.K.M. and S.M. designed study, analyzed data and wrote manuscript.


\appendix

\section{Stochastic load sharing}

\begin{figure}[h!]
	\centering
	\includegraphics[width=0.9\linewidth]{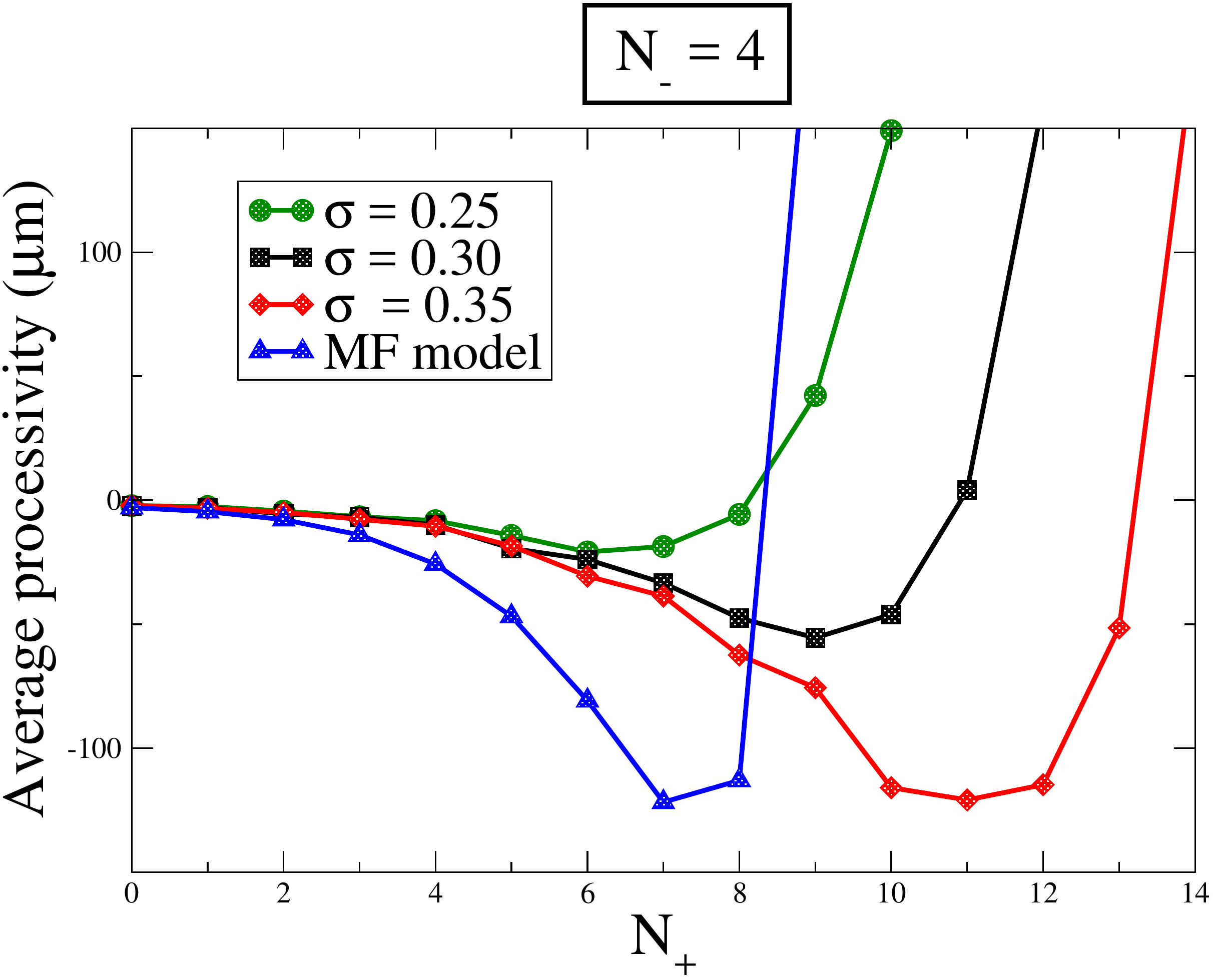}
	\caption{Average processivity as a function of $N_+$, as the bead size $\sigma$ is changed. Note that the friction constant $\zeta$ changes as a result. The blue curve shows the corresponding result under the equal load sharing assumption. Here $N_- = 4$, $\alpha = 40$ and $F_o = 7pN$.}
	\label{fig:Sbd}
\end{figure}

	In order to ensure that the codependent transport characteristics obtained are not artifacts of the mean field assumption, where motors are assumed to share the load force equally, we also performed Brownian dynamic simulations where the load is shared stochastically, with each motor having a different extension, and hence facing a different opposing load.
	
	In the simulation, $N$ motors are attached to the cargo. The motors are modeled as elastic springs with spring constant $k = 0.32$ pN/nm. The springs have a rest length $l_0$ and generate a restoring force only when stretched beyond the rest length. The rest length of the springs are chosen in accordance with earlier simulations, $l_0 = 100$ nm for kinesin and $l_0 = 50$ nm for dynein. In this one dimensional model, we start by putting the bead at the origin and all $N$ motors attached irreversibly to the cargo at one end. The other end of the motors are allowed to bind to any point on the track within the rest length of the corresponding motor, on either side of the bead. 
	 
	 At every time step, all the $N$ motors are visited to determine if they are in the attached or detached state. Each motor position and their state are updated only once in a time step. If the motor is in the detached state, then it can re attach with a probability $P_{\textrm{on}} = \pi_{\pm} \Delta t$, where $\pi_{\pm}$ are the binding rates of kinesin and dynein as defined earlier. The attachment happens within a distance $l_0$ on either side of the bead.  If the $i$th motor is in an attached state, then the load force, $F_i$ is calculated by multiplying the extension of the spring with the spring constant $k$. Depending on the load force, the motor could detach, with probability $P_{\textrm{off}} = \varepsilon_{\pm}(F_i) \delta t$, where $\varepsilon_{\pm}$ are the unbinding rates of kinesin and dynein. Note that for dynein, $F_i$ replaces $F_c$ in Eq. 2 in the manuscript. If the motor does not detach, then we calculate the probability of taking a step, $P_{\textrm{step}} = k_{\textrm{step}} \Delta t$, where $k_{\textrm{step}} = (v_{0\pm}/d)(1 - F_i/F_{s\pm})$, where $v_{0\pm}$ is the unloaded velocity of the single motor, $F_{s\pm}$ is the stall force of the motor and $d = 8$ nm is the step length of the motor. Note that this form is used for backward loads $F_i < F_s$. For backward loads $F_i > F_s$, $P_{\textrm{step}} = 0$. For forward loads, $F_i = 0$. If the motor steps, its position is updated from $x_i$ to $x_i + d$. All motor states and their positions are updated simultaneously in a given time step. Two sets of motors with their characteristic parameters as given in Table~\ref{tab1}, move in opposite directions. 	 
	 
	 To update the position of the cargo (modeled as a bead of radius $\sigma$), we calculate the total force acting on the cargo due to both sets of molecular motors moving in opposite directions, $F_{\textrm{tot}} = \sum F_i$. Note that the detached motors do not contribute to the total force, neither do the motors which lie within a rest length from the bead position. The bead is under the influence of both thermal and viscous forces with $\xi = 0.001$ pN-s/$\mu-m^2$ being the viscosity of the medium. The bead diffuses with diffusion constant $D = k_BT/\zeta$ where $\zeta = 6\pi\xi \sigma$ is the friction constant. When the cargo is subjected to the force $F_{\textrm{tot}}$ it moves with the velocity $v_d = F_{\textrm{tot}}/\zeta$. In the presence of thermal noise, the overdamped Brownian dynamics of the cargo is given by  
	 
	 \begin{equation}
	 {\bf x}(t + \Delta t)  = {\bf x} (t) + v_d\Delta t + \eta 
	 \end{equation}
	 
	 where $\eta$  are drawn Gaussian distribution with $\langle \eta(t) \rangle = 0$ and $\langle \eta(t) \eta(t^{\prime})\rangle = 2D \delta(t - t^{\prime})$.

\section{Fokker Planck equation}

	If the maximum number of kinesin ($N_+$) and dynein ($N_-$) is large, $N_+, N_- \gg 1$, we can expand the probabilities in the Master equation in a Taylor series to obtain the associated Fokker Planck equation. We define,
	$x=n_+/N_+$ and $y=n_-/N_-$, and in terms of these variables,
	\begin{eqnarray}
	p(x\pm\frac{1}{N_+},y) &=& p(x,y) \pm \frac{1}{N_+} \partial_x p(x,y) + \frac{1}{2 N_+^2} \partial_x^2 p(x,y) \nonumber \\
	p(x,y\pm\frac{1}{N_-}) &=& p(x,y) \pm \frac{1}{N_-} \partial_y p(x,y) + \frac{1}{2 N_-^2} \partial_y^2 p(x,y) \nonumber \\
	\epsilon_+(x+\frac{1}{N_+},y) &=& \epsilon_+(x,y) + \frac{1}{N_+} \partial_x \epsilon(x,y) + \frac{1}{2 N_+^2} \partial_x^2 \epsilon(x,y) \nonumber \\
	\epsilon_-(x,y+\frac{1}{N_-}) &=& \epsilon_-(x,y) + \frac{1}{N_-} \partial_y \epsilon(x,y) + \frac{1}{2 N_-^2} \partial_y^2 \epsilon(x,y) \nonumber \\
	\pi_+(x-\frac{1}{N_+},y) &=& \pi_+(x,y) - \frac{1}{N_+} \partial_x \pi(x,y) + \frac{1}{2 N_+^2} \partial_x^2 \pi(x,y) \nonumber \\
	\pi_-(x,y-\frac{1}{N_-}) &=& \pi_-(x,y) - \frac{1}{N_-} \partial_y \pi(x,y) + \frac{1}{2 N_-^2} \partial_y^2 \epsilon(x,y)\nonumber\\
	\end{eqnarray}
	Substituting in the Master equation, Eq.~A1, and neglecting terms of order $\mathcal{O}(1/N_{\pm}^3)$, we obtain,
	\begin{eqnarray}
	\frac{\partial}{\partial t} p(x,y,t) = -\sum_{1}^{2} \frac{\partial}{\partial {x_i}} \left [ v_i(x,y) p(x,y,t) \right ] + \nonumber \\ \sum_{i=1}^{2} \sum_{j=1}^{2} \frac{\partial^2}{\partial x_i \partial x_j} \left [ D_{ij} (x,y) p(x,y,t) \right ] 
	\end{eqnarray}
	where, 
	\begin{eqnarray}
	v_x (x,y) &=& \frac{1}{N_+} \left (\pi_+(x,y) - \epsilon_+(x,y) \right ) \nonumber \\
	v_y (x,y) &=& \frac{1}{N_-} \left (\pi_-(x,y) - \epsilon_-(x,y) \right ) \nonumber \\
	D_{xx} (x,y) &=& \frac{1}{2 N_+^2} \left (\pi_+(x,y) + \epsilon_+(x,y) \right ) \nonumber \\
	D_{yy} (x,y) &=& \frac{1}{2 N_-^2} \left (\pi_-(x,y) + \epsilon_-(x,y) \right ) \nonumber \\
	D_{xy} (x,y) &=& D_{yx} (x,y) ~=~ 0 
	\end{eqnarray}
	
	The analysis of the FPE and the comparison of the steady state probabilities with those obtained by the numerical solution of the Master Equation will be presented in a separate manuscript.

\section{Average Processivity}

\begin{figure*}[t]
	\centering
	\includegraphics[width=\linewidth]{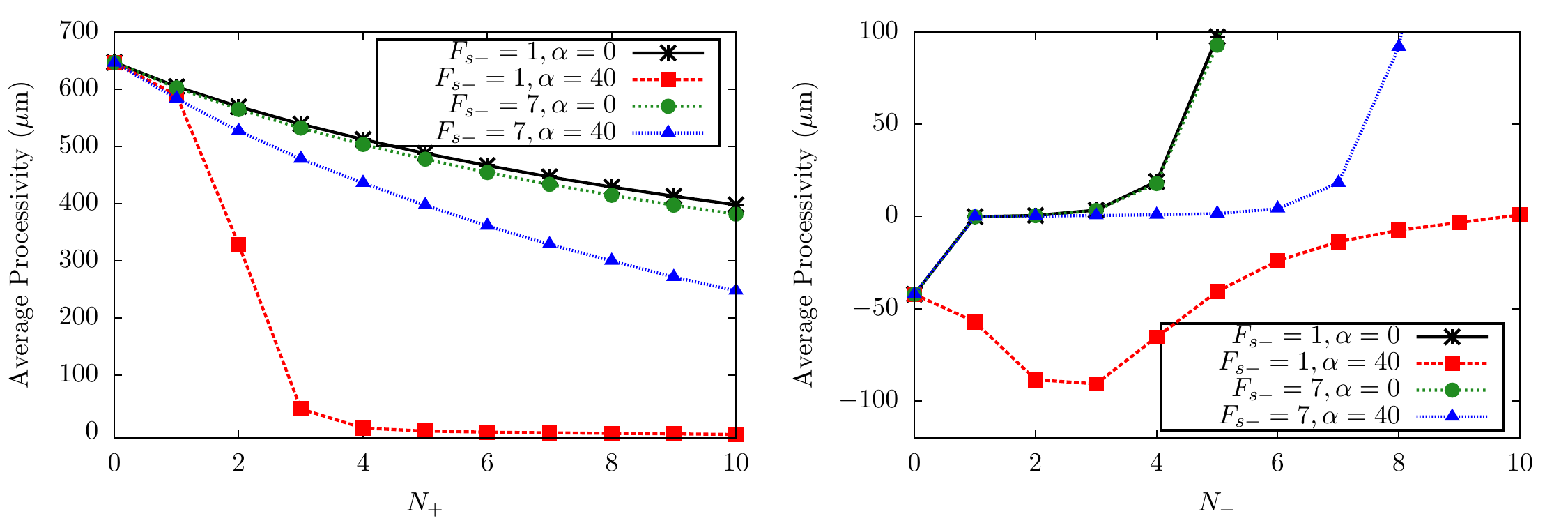}
	\caption{Average processivity (a) as a function of $N_-$ for $N_+ = 9$ , and (b) as a function of $N_+$ for $N_- = 9$. The zero-force (un)binding rates for dynein are $\varepsilon_{0-} = \pi_{0-} = 1/s$, and $F_o = 7pN$.}
	\label{fig:S1}
\end{figure*}
\begin{figure*}[t]
	\centering
	\includegraphics[width=\linewidth]{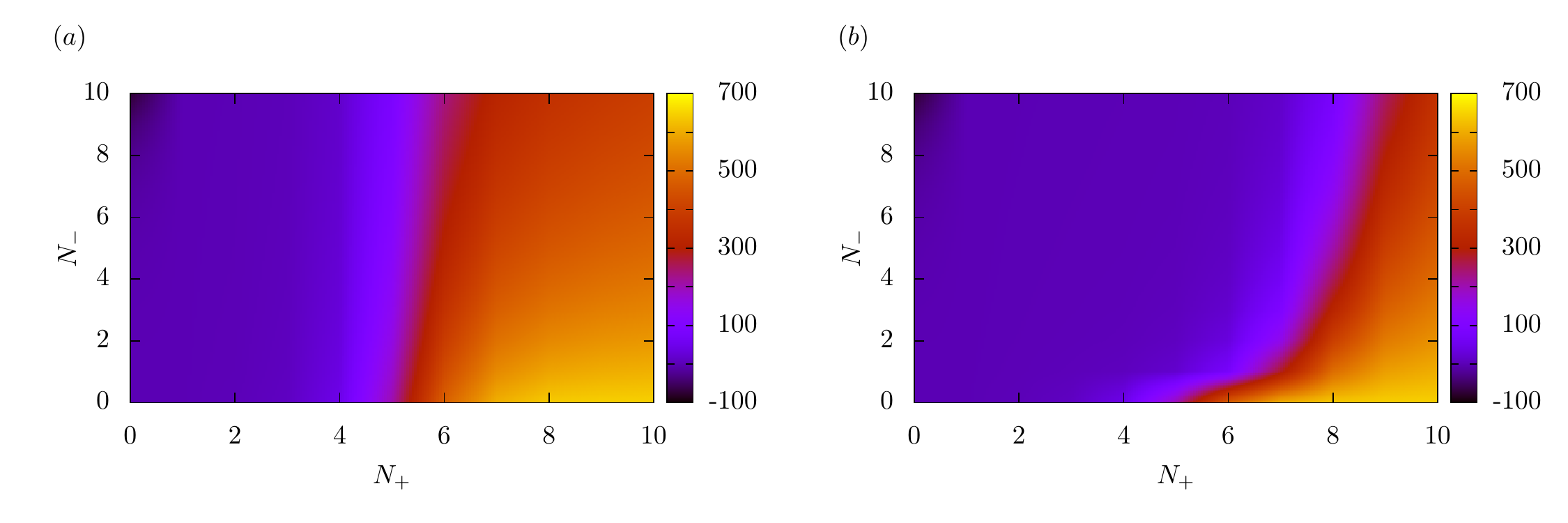}
	\caption{Processivity contour plots in the $N_+ - N_-$ plane for strong dynein (a) without catch bond ($\alpha = 0$); and (b) with catch bonds ($\alpha = 40$,  $F_o = 7pN$). The colorbar indicates the average processivity (in $\mu m$). Yellow regions denote strong plus ended runs, while dark blue regions indicate strong minus ended runs. The zero-force (un)binding rates for dynein are $\varepsilon_{0-} = \pi_{0-} = 1/s$ }
	\label{fig:S2}
\end{figure*}

In Fig.~\ref{fig:S1}, we look at the variation of the average processivity as (a) a function of $N_-$ for fixed $N_+ = 9$ (Fig.~\ref{fig:S1}a), and (b) a function of $N_+$ for fixed $N_- = 9$ (Fig.~\ref{fig:S1}b). In Fig.~\ref{fig:S1}(a), we find that the average processivity decreases with $N_-$ for non-catchbonded dynein. Catch-bonded dynein with strong tenacity exhibits qualitatively similar behaviour as that of motor without catchbond with the processivity decreasing on increasing $N_-$ , while dyneins with weaker tenacity can stall the motion of the cargo. This again arises because the catch bond is activated at smaller opposing loads for weak dynein, leads to drastic effects on the motion of the cargo.

In Fig.\ref{fig:S1}(b), we show that again, while strong dynein exhibits qualitatively similar behaviour to non-catch bonded dynein, weak dyneins show a counter-intuitive codependent behaviour as was seen in Fig. 2(b). As the number of kinesin motors increases initially, the cargo walks more in the negative direction, with the cargo walking an average $\sim 100 \mu m$ in the negative direction for $2-3$ kinesin molecules, compared to around $\sim 40 \mu m$ when no kinesins are present. Beyond 3 kinesin motors, on increasing the kinesin number, the processivity in the negative direction increases, as would be expected from the normal mechanical tug-of-war picture.

In Fig.~\ref{fig:S2} we show the contour plots of the processivity of the cargo for strong dynein in the $N_+ - N_-$ plane. In the absence of catch bond, the contour plots looks similar to the one for weak dynein (Fig. 2c), with strong positive directed runs for $N_+ > 5$. Strong negative runs are achieved only for very high dynein number coupled to very low kinesin number. In the presence of the catch bond, dyneins are able to counteract the positive directed load more efficiently, with strong positive runs occurring for higher number of kinesins than in the non-catch bonded case. The special cases corresponding to $N_+ / N_- = 4$ are shown in Fig. 2(a) and (b), while the case corresponding to $N_+ / N_- = 9$ are shown in Fig.~\ref{fig:S1}(a) and (b).

\section{Average number of bound motors}

  In Fig.~\ref{fig:Sbd}, we look at the average processivity as the friction constant $\zeta$ is varied by varying $\sigma$. The rest of the parameters are as in fig. 2(b) of the main text with $\alpha = 40$ and  $F_o = 7pN$. For all three values of $\sigma$, the non-monotonic behavior is reproduced indicative of the catchbonded codependent behavior. With increasing $\sigma$, the cusp is observed at higher values of $N_+$ with average processivities close to values obtained in our mean field model. 

\begin{figure*}[t]
	\centering
	\includegraphics[width=0.8\linewidth]{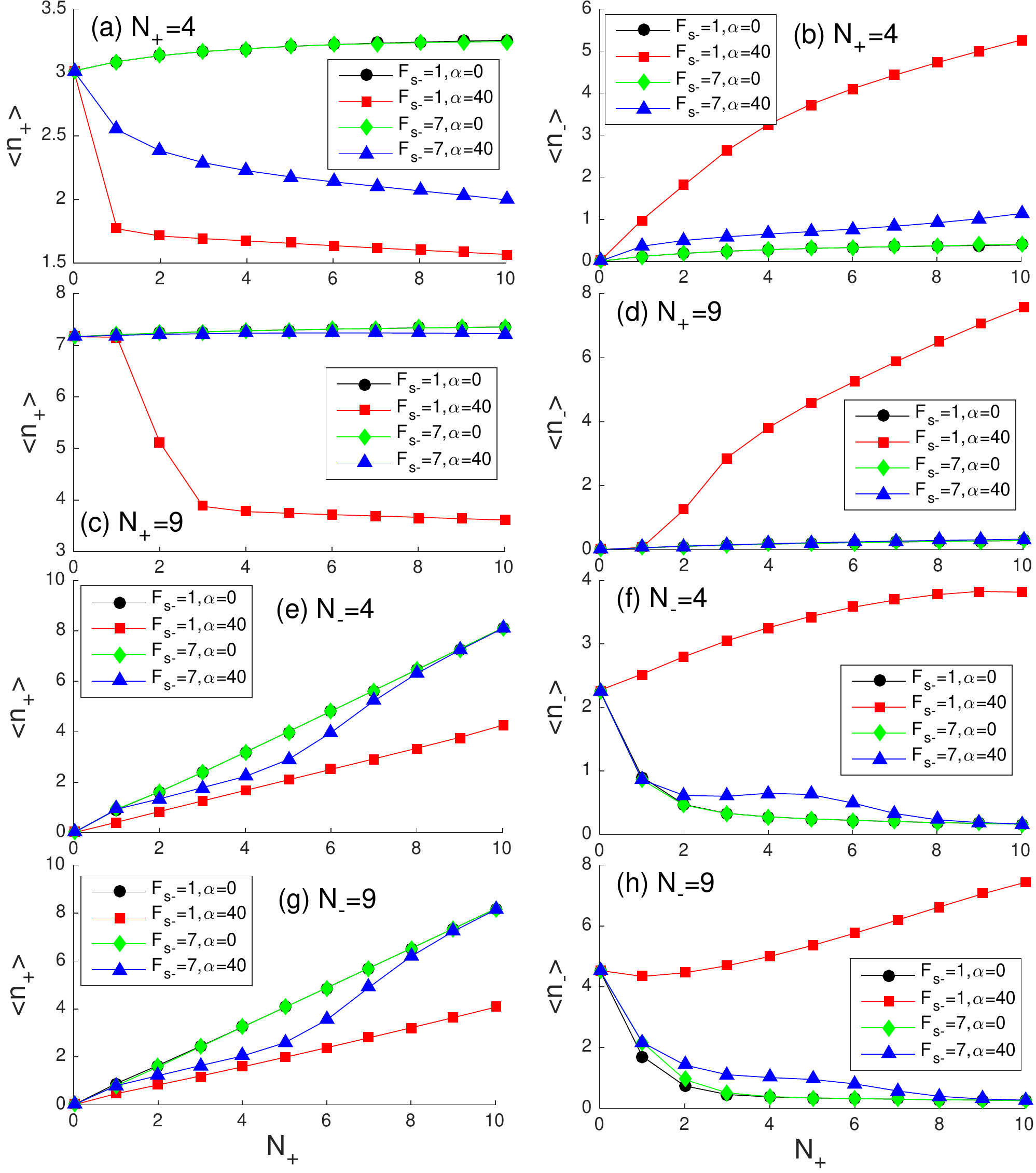}
	\caption{Average number of bound kinesins for (a) $N_+ = 4$, (c) $N_+ = 9$, (e) $N_- = 4$, (g) $N_- = 9$. Average number of bound dyneins for (b) $N_+ = 4$, (d) $N_+ = 9$, (f) $N_- = 4$, (h) $N_- = 9$. The data for both strong and weak dynein with and without catch bond. Here  $F_o = 7pN$.}
	\label{fig:S3}
\end{figure*}

The effect of catchbonding on the processivity can also be understood in terms of the average number of bound motors. As illustrated in Fig.~\ref{fig:S3},  in the absence of catch bond, the number of attached dyneins shows a very weak increase with increasing $N_-$, saturating at a value of $\sim 0.3$. The average number of kinesins is roughly around 3, leading to strong positive runs in the absence of catch bonds. For catch bonded weak dynein $F_{s-} = 1 pN$, on increasing $N_-$, the average number of bound dyneins increases sharply. For low $N_-$, the average number is almost the same as the maximum number, $\langle n- \rangle \sim N_-$. The average number of attached kinesins also falls sharply to under $2$. On increasing $N_-$ even further, the average number of bound dyneins keeps increasing, while the average number of bound kinesins roughly remains constant. The higher number of bound dyneins lead to overall minus directed runs in this regime. For strong dynein, both the increase in $\langle n- \rangle$ and $\langle n_+ \rangle$ are much less sharp, illustrating that catch bond plays a less drastic role here in contrast to weak dynein. The average number of bound kinesins and dyneins are comparable in this case, which effectively leads to no net motion for $N_- \geq 2$.

The behaviour of the processivity as a function of $N_+$ can also be understood in terms of the average number of bound motors.  As shown in Fig.~\ref{fig:S3}, in the absence of catch bonds, on increasing $N_+$, the average number of dyneins fall drastically, approaching $\sim 0.1$ for large values of $N_+$. In contrast, the average number of bound kinesins increases linearly, leading to stronger plus-end directed runs with increasing $N_+$. For catch bonded weak dynein ($F_{s-} = 1 pN$), remarkably, the average number of attached dyneins {\it increases} with increasing $N_+$. This is again a direct consequence of the catch bond, where the increasing opposing force due to more kinesin motors pushes dyneins into the catch bonded state, effectively increasing their numbers. This leads to the increased processivity of the cargo in the negative direction with increasing $N_+$ as shown in Fig. 2(b). The effect for $F_{s-} = 7 pN$ dynein is much more muted, since it is difficult to push these strong dyneins into the catch bonded regime.

\section{Catchbond Strength}
\begin{figure}[h!]
	\centering
	\includegraphics[width=\linewidth]{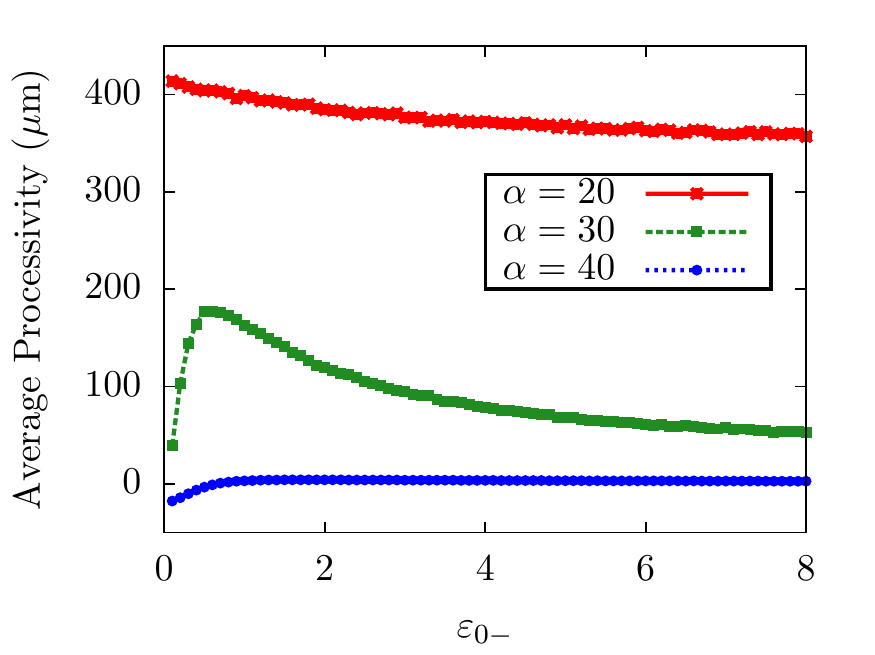}
	\caption{Processivity as a function of $\varepsilon_{0-}$ for different catch bond strengths ($\alpha$) for weak dynein ($F_{s-} = 1 pN$). Data shown is for $N_+ = 6$, $N_- = 2$,  $\pi_{0-} = 1/s$ and $F_o = 7pN$.}
	\label{fig:S44}
\end{figure}
The strength of the dynein catch bond ($\alpha$) is a phenomenological parameter in our model. Changing the strength of the catch bond can have dramatic consequences for the processivity characteristics. This is shown in Fig.~\ref{fig:S44} for three values of the catch bond strength. For $\alpha = 20 k_BT$, on increasing the dynein unbinding rate, the average processivity in the positive direction decreases monotonically. For $\alpha = 30 k_B T$, on weakening the dynein, the processivity in the positive direction initially increases, as expected from standard tug-of-war. However, beyond a certain $\varepsilon_{0-}$, weakening the dynein further, causes a net decreases in the processivity in the positive direction. Finally, for $\alpha = 40 k_B T$, on increasing $\varepsilon_{0-}$, the runlength in the negative direction initially decreases, and beyond a certain point, saturates to almost zero, becoming insensitive to further changes in the unbinding rate, as has been discussed in the main text for Fig.~3(a). 

\begin{figure*}[t]
	\centering
	\includegraphics[width=\linewidth]{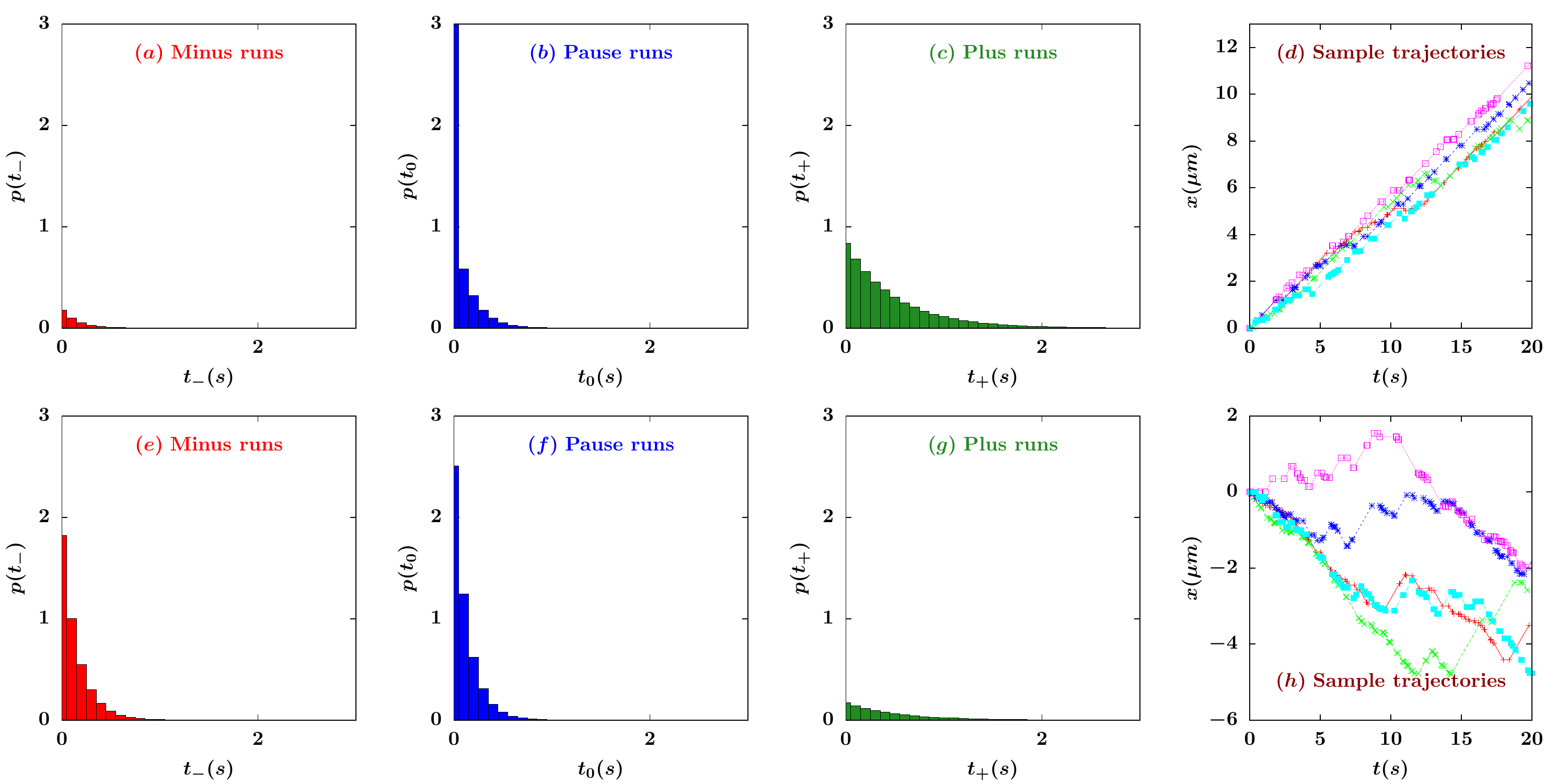}
	\caption{Probability distributions of runtime for $N_+ = 1$, $N_- = 1$. The top panel shows the normalized histograms and sample trajectories for dynein in the absence of catch bond ($\alpha = 0$). The bottom panel shows the corresponding quantities for catch-bonded dynein ($\alpha = 40$, $F_o = 7pN$). (a) and (e) Distributions of runtime for minus directed runs (shown in red); (b) and (f) pausetime distributions (shown in blue); (c) and (g) distributions of runtime for plus directed runs (shown in green); and  (d) and (h) sample trajectories. Insets, where present, show a magnified view of the probability distributions.}
	\label{fig:S4}
\end{figure*}

\section{Probability distributions and sample trajectories}

Here, we analyse the case of bidirectional cargo transport by 1 kinesin motor and 1 dynein motor. In Fig.~\ref{fig:S4}(a), (b) (c)  we display the distribution of runtime along the negative direction, the distribution of times the cargo spends in pause state (which arises due to the  simultaneous attachment of dynein and kinesin motors to the filament leading to {\it tug-of-war}), and the distribution of runtime along the positive direction, in the absence of catchbonding in dynein.  In this scenario, the frequency of the positive runs exceeds the negative runs by almost one order of magnitude as indicated by looking at the peaks of the probability distribution of the run time. This can be understood as a direct consequence of the relatively high tenacity of the kinesin with respect to dynein. Since stall force of the kinesin motor is around 5 times that of dynein, in a typical situation of tug-of-war, in the absence of catchbonding, the unbinding rate of the dynein motors due to the opposing load of the kinesin motors rises far steeper than that of the unbinding rate of kinesin motor due  to the opposing load of dynein motors. This leads to preponderance of the positive runs viz-a-viz negative runs. Further, the average runtime along the positive directions is more than the run lengths along the negative direction. This is simply a consequence of the fact that the kinesin binding rates are chosen to be higher than the dynein binding rates (see Table 1). Thus a plus-moving run, on average, continues for a longer time than a minus run, leading to larger average runtime along the positive direction. This trivially implies that the runlengths in the positive direction are also larger than those along the negative direction, the runlength being related linearly to the runtime through the forward velocity of the motor ($v_{F+} = v_{F-} = 0.65 \mu m /s$). As seen in Fig.~\ref{fig:S4}(b), in the absence of catchbond the average time the cargo spends in the paused state is an order of magnitude smaller than the time it spends in the plus moving state. Thus overall the motion of the cargo in the absence of catchbond for this case is strong plus ended motion with weak pauses and negligible runs along the negative direction ( Fig.~\ref{fig:S4}d). Incorporation of the catchbonding behaviour of dynein is demonstrated by comparison of the probability distributions of runlengths and pauses as depicted  in Fig.~\ref{fig:S4}(e), (f) (g) with Fig.~\ref{fig:S4}(a), (b), (c) . First of all  by comparing Fig.~\ref{fig:S4}(e) and 4(g) it can be seen that the frequency of the negative runs now exceed that of the positive directed runs. This is due to the fact that in the tug-of-war state, when the attached dynein experiences the load force due to the attached kinesin, the dynein enters a catchbonded state and thus its propensity to unbind from the filament diminishes, while that of the kinesin remains unaltered, resulting in a situation where the pause state is more often transformed into a minus-end directed state of the cargo. While the characteristic pause times do not change substantially, they now become comparable to the runtime in the negative direction, as can be seen by comparing Fig.~\ref{fig:S4}(f) with Figs.~\ref{fig:S4}(b) and (e). The average run length either along the plus or the minus end remains unaffected due to catchbonding with the average runs along the plus direction being higher than that of the dynein due to higher kinesin binding rates compared to the dynein motors.  The corresponding trajectories ( Fig.~\ref{fig:S4}(g) ) then correspond to bidirectional motion, characterized by more frequency of negative runs but longer average plus ended runs, and more prominent pauses. 

\bibliographystyle{prsty}

\end{document}